\RequirePackage{etoolbox}
\documentclass[aoas,preprint]{imsart}
\pdfoutput=1

\usepackage{amsthm}
\usepackage{amsmath}
\usepackage{natbib}
\usepackage{graphicx}

\usepackage{epstopdf}
\usepackage{wrapfig}
\usepackage{lscape}
\usepackage[figuresright]{rotating}
\usepackage{subfigure}
\usepackage{float}

\let\ORGsidewaysfigure\sidewaysfigure
\let\ORGendsidewaysfigure\endsidewaysfigure
\renewenvironment{sidewaysfigure}
  {\ORGsidewaysfigure\vspace*{.5\textwidth}\begin{minipage}{\textheight}\centering}
  {\end{minipage}\ORGendsidewaysfigure}

\usepackage{epstopdf}
\newcommand{\E}{\mbox{E}}
\newcommand{\X}{\mbox{X}}
\newcommand{\mX}{\mathbf{X}}

\newcommand{\Y}{\mbox{Y}}

\newcommand{\N}{\mbox{N}}

\newcommand{\appropto}{\mathrel{\vcenter{
  \offinterlineskip\halign{\hfil$##$\cr
    \propto\cr\noalign{\kern2pt}\sim\cr\noalign{\kern-2pt}}}}}

\startlocaldefs
\numberwithin{equation}{section}
\theoremstyle{plain}

\endlocaldefs

\begin{document}

\begin{frontmatter}
\runtitle{Regularized treatment effect regression}

\begin{aug}
\author{\fnms{P. Richard} \snm{Hahn}\ead[label=e1]{richard.hahn@chicagobooth.edu}},
\author{\fnms{Carlos M.} \snm{Carvalho}\ead[label=e2]{carlos.carvalho@mccombs.utexas.edu}},\\
\author{\fnms{Jingyu} \snm{He}\ead[label=e3]{jingyuhe@chicagobooth.edu}}
\and
\author{\fnms{David} \snm{Puelz}\ead[label=e4]{david.puelz@utexas.edu}}\\
\vspace{0.1in}
\title{Regularization and confounding \\in linear regression for\\ treatment effect estimation}

\runauthor{Hahn, Carvalho, He and Puelz}

\address[addr1]{\printead{e1}}
\address[addr2]{\printead{e2}}
\address[addr3]{\printead{e3}}
\address[addr4]{\printead{e4}}

\end{aug}

\begin{abstract}
This paper investigates the use of regularization priors in the context of treatment effect estimation using observational data where the number of control variables is large relative to the number of observations.  First, the phenomenon of ``regularization-induced confounding'' is introduced, which refers to the tendency of regularization priors to adversely bias treatment effect estimates by over-shrinking control variable regression coefficients. Then, a simultaneous regression model is presented which permits regularization priors to be specified in a way that avoids this unintentional ``re-confounding". The new model is illustrated on synthetic and empirical data. 
\end{abstract}


\end{frontmatter}

\section{Introduction}
This paper considers the use of Bayesian regularized  linear regression models for the purpose of estimating a treatment effect from observational data. Treatment effects ---  the amount some response variable would change if the value of the treatment variable were changed by a given amount --- can only be properly estimated from observational data by taking into account all of the various explanatory factors that may otherwise account for the observed correlation between the treatment and response variables. In the case of a linear regression model (assuming it to be correct) this ``adjustment for confounding'' means that the model includes a sufficient set of control variables as regressors in addition to the treatment variable.


Practical implementation of regression modeling for estimating treatment effects from observational data is complicated by two related issues. First, the minimal set of sufficient control variables is almost never known and second, the set of candidate control variables is often quite large relative to the available sample size. This consideration suggests that statistical regularization has a role to play in reliable treatment effect estimation. It may therefore come as a surprise that naive deployment of Bayesian shrinkage priors in the context of treatment effect estimation can yield exceptionally poor estimators. Exploring this phenomenon and providing a straightforward solution is the main contribution of this paper. We show that regularization can indeed provide statistical improvements over maximum likelihood estimation, but that it must be imposed carefully, in a sense we will make precise.


\subsection{Previous literature}
Treatment effect estimation is an important topic with a long and varied literature; a comprehensive review is beyond the scope of this paper. For review articles from an expressly Bayesian perspective, see \cite{li2014bayesian} or \cite{heckman2014treatment}. This paper focuses more narrowly on the impact of regularization or ``shrinkage" priors on the estimation of treatment effects from observational studies. Our use of regularization priors in this context addresses a practical data analysis problem that has been recognized since at least \cite{leamer1983let}: regression analyses including very many potential control variables often produce unsatisfyingly imprecise effect estimates. \cite{leamer1983let} admonishes those who react to this dilemma by hand-selecting a small subset of the potential controls and proceeding with analysis as if the others were irrelevant. See also \cite{leamer1978specification} for an early Bayesian treatment of this problem.


More specifically, this paper represents a contribution to the small but growing literature on Bayesian approaches to treatment effect estimation via linear regression with many potential controls. Specifically, we propose a conceptual and computational refinement of ideas first explored in \cite{wang2012bayesian}, where Bayesian adjustment for confounding is addressed via hierarchical priors. Our proposed method can be seen as an alternative to \cite{wang2012bayesian}, with certain conceptual and computational advantages, namely ease of prior specification and posterior sampling. Other papers elaborating upon this approach include \cite{wang2015accounting}, \cite{lefebvre2014effect} and \cite{talbot2015bayesian}; see also \cite{jacobi2016bayesian}. \cite{zigler2014uncertainty} and \cite{an2010bayesian} focus on Bayesian propensity score models (for use with binary treatment variables). \cite{wilson2014confounder} takes a decision theoretic approach to variable selection of controls.  Again, each of these previous approaches cast the problem as one of selecting appropriate controls; posterior treatment effect estimates are obtained via model averaging.  Here, we argue that if the goal is estimation of a certain regression parameter (corresponding to the treatment effect, provided the model is correctly specified), then questions about which specific variables are necessary controls is a means to an end rather than an end in itself. Other recent papers looking at regularized regression for treatment effect estimation include \cite{ertefaie2015variable} and \cite{ghosh2015penalized}, but even here the focus is on variable selection via the use of $1$-norm penalties on the regression coefficients.

Finally, treatment effect estimation is clearly a sub-topic within the broader field of causal inference. Here, we do not emphasize this connection, focusing instead on the specifics of the important special case that is linear regression. For excellent book length treatments on causal inference, we recommend  \cite{imbens2015causal} and \cite{morgan2014counterfactuals}. Like \cite{wang2012bayesian}, our work has forebears in earlier work based on joint modeling of treatment and response variables as functions of control variables, notably \cite{rosenbaum1983central} and \cite{robins1992estimating}, as well as \cite{mccandless2009bayesian}.

\subsection{Outline}
The paper is structured as follows. In section \ref{reparametrization}, we describe how naive regularization can corrupt treatment effect estimation and present a reparametrized linear model that avoids this pitfall. Section \ref{simulations} presents extensive simulation studies demonstrating the performance of the new model relative to standard alternatives. Section \ref{levitt} reanalyses the data of \cite{donohue2001impact}, which considers the impact of abortion laws on crime rates, following the similar recent (frequentist) analysis of \cite{belloni2014inference}.

\section{Regularized linear regression for treatment effect estimation}\label{reparametrization}
In this paper, we focus on linear regression models
\begin{equation}\label{responseeqn}
	Y_i = \alpha Z_i + \X_i \beta + \nu_i,
\end{equation} where $\X_{i}$ is a row vector of control variables, $\beta$ is a column vector of the control effects, $Z_{i}$ is a continuous scalar treatment variable and $\alpha$ is a scalar regression coefficient. When these variables are meant to be interpreted as random variables, they will be denoted in capital letters; when they are to be interpreted as observed quantities they will either be lower case, to indicate a scalar quantity, roman font, to indicate a vector, or bold, to indicate a matrix.  We assume the errors, $\nu_{i}$, are normally distributed with zero mean and unknown variance. Under these assumptions, the ordinary least squares estimator gives unbiased estimates with valid coverage. 

Our goal is to accurately estimate the treatment effect, and this is done by including the proper controls in the equation. Specifically, ``proper" in this context means that: 
\begin{equation}\label{covcond}
	\mbox{cov}(Z_{i}, \nu_{i} \vert \X_{i}) = 0.	
\end{equation}This {\em exogeneity} condition guarantees that estimates of $\alpha$ will have the desired counterfactual interpretation as ``the amount $Y$ {\em would change} if $Z$ {\em were changed} by one unit'': $\alpha = \E(Y \mid Z = z+1, \X) - \E(Y \mid Z = z, X)$. For a detailed discussion of why (\ref{covcond}) licenses a causal interpretation, see e.g. \cite{imbens2015causal} section 12.2.4 or \cite{morgan2014counterfactuals} section 6.2. 

It will be assumed throughout that this model is correctly specified so that attention may be focused narrowly on the impact that regularization has on posterior inferences regarding parameter $\alpha$. To emphasize, a thorough regression analysis for causal inference should including a sensitivity analysis to gauge robustness of one's inferences to various modeling assumptions. In this paper we intentionally set these practically important concerns aside for conceptual clarity: the phenomenon of ``regularization-induced confounding'' is an independent issue that arises even if the model and exogeneity assumptions are all satisfied. For a complete introduction to the host of additional issues surrounding causal inference, see again \cite{imbens2015causal}.

The most common approach to estimating the parameters of linear regression models is via ordinary least squares (OLS), which in the present model is equivalent to maximum likelihood estimation. Estimating the model via OLS guarantees that (\ref{covcond}) is satisfied by construction: it is well-known, and easy to show, that the residual vector produced by OLS has zero correlation with the observed treatment vector $Z$. However, (\ref{covcond}) will not in general be satisfied by the residuals corresponding to a shrinkage estimator of $\beta$. Accordingly, in finite samples we have two competing criteria -- the shrinkage prior over $\beta$ and the sampling distribution for $Y_{1:n}$ -- which combine to form our eventual estimate. What can happen in this setting is that posterior inferences can be affected by the prior in such a way that (\ref{covcond}) is violated in-sample, making the causal interpretation of the $\alpha$ estimate suspect. 

Intuitively, the prior ``prefers'' to have ``small'' elements of $\beta$; in the case of strong confounding, a very similar in-sample fit can be achieved by over-stating the magnitude of the treatment effect parameter $\alpha$ (which is one-dimensional) while simultaneously attenuating the control variable coefficients. To observe this phenomenon formally, we can examine the bias of the posterior mean of $\alpha$ in the case of a standard normal (ridge) prior over $\beta$. In this case, considering $\mathrm{z}$ and $\mathbf{X}$ fixed, the bias of $\hat{\alpha}_{rr}$ ($rr$ for ``ridge regression'') under an independent non-informative prior, may be expressed as:
\begin{equation}\label{bias}
\mbox{bias}(\hat{\alpha}_{rr}) = -\left((\mathrm{z}^t\mathrm{z})^{-1}\mathrm{z}^t\mathbf{X}\right)(\mathbf{I}_p + \mathbf{X}^t(\mathbf{X} - \hat{\mathbf{X}}_Z))^{-1}\beta.
\end{equation}
The first term is a $p$-vector of regression coefficients corresponding to univariate regressions of each $X_j$ on $Z$; $\hat{\mathbf{X}}_Z$ is the $n$-by-$p$ matrix of fitted values from these $p$ regressions. Note that the bias is not a function of the true value of $\alpha$, but is a function of every element of the true (unknown) $\beta$ vector, with weights proportional to how well $X_j$ is predicted by $Z$. To put this formula into the context of treatment effect estimation, it says roughly that the stronger the confounding is, the worse the bias on the treatment effect parameter $\alpha$ will be.

%



\subsection{A reparametrized model for regularized treatment effect regressions}
Consider the two equation model:
\begin{equation} \label{themodel0}
	\begin{split}
	\text{Selection Eq.:} &\hspace{5mm}	Z = \X\gamma + \epsilon, \hspace{15.5mm} \epsilon \sim N(0,\sigma_{\epsilon}^{2}), 
		\\
		\text{Response Eq.:} &\hspace{5mm}	Y = \alpha Z + \X\beta + \nu, \hspace{5mm} \nu \sim N(0,\sigma_{\nu}^{2}).
	\end{split}
\end{equation}
Without loss of generality, assume that our variables are zero centered (in practice, one may include an intercept term).

The designation ``selection" refers to the impact that the control variables have on the level of treatment, $Z$, received. Prototypically, certain individuals are ``selected" to receive treatment. The ``response" equation describes the impact of the treatment and controls, $\X$, on the conditional expectation of the response (outcome) variable, $Y$. Prototypically, $Y$ records some diagnostic measure on individuals. Because $\X$ appears in both equations, the selection equation reflects the confounding influence of the controls, and the residual variance of this equation, $\sigma_{\epsilon}$, gauges the extent of the confounding. 

These equations correspond to the factorization of the joint distribution $$f(Y, Z \mid X, \gamma, \beta, \sigma_{\epsilon}, \sigma_{\nu}) = f(Y \mid X, Z, \beta, \sigma_{\epsilon}, X)f(Z \mid X, \gamma, \sigma_{\nu},).$$ This factorization implies a complete separation of the parameter sets; specifically, independent priors on the regression parameters $\pi(\beta, \gamma, \alpha) = \pi(\beta)\pi(\gamma)\pi(\alpha)$ imply that only the response equation is used in estimating $\beta$ and $\alpha$. 

It is possible, as investigated in \cite{wang2012bayesian}, to incorporate information concerning $\gamma$ into the inference for $\beta$ via a joint prior $\pi(\beta \mid \gamma)\pi(\gamma)$ which would then be updated by the treatment data as $\pi(\gamma \mid Z)$ whereupon it can be incorporated with the response likelihood via the integrated prior $\pi(\beta \mid Z) = \int_{\gamma} \pi(\beta \mid \gamma)\pi(\gamma \mid Z)$. Our approach will be more direct, placing widely-used independent priors in a transformed parameter space.

Specifically, we introduce the following transformation:
\begin{equation}\label{transformation}
\begin{pmatrix}
\alpha\\
\beta +  \alpha \gamma\\
\gamma
\end{pmatrix} \rightarrow \begin{pmatrix}
\alpha \\
\beta_d\\
\beta_c
\end{pmatrix},
\end{equation}
which yields the model
\begin{equation} \label{themodel}
	\begin{split}
	\text{Selection Eq.:} &\hspace{5mm}	Z = \X\beta_c + \epsilon, \hspace{31mm} \epsilon \sim N(0,\sigma_{\epsilon}^{2}), 
		\\
		\text{Response Eq.:} &\hspace{5mm}	Y = \alpha(Z - \X\beta_c) + \X\beta_d + \nu, \hspace{5mm} \nu \sim N(0,\sigma_{\nu}^{2}).
	\end{split}
\end{equation}

Our approach will be to place  independent regularization priors over $\beta_c$ and $\beta_d$ and to update our prior using the likelihood of both of the above equations. 

This parametrization tidily separates the distinct roles that covariates can play in a regression analysis of causal effects. Specifically, in previous literature, a ``prognostic'' or ``predictive'' variable refers to variables $X_j$ with $\beta_{c,j} = 0$, $\beta_{d,j} \neq 0$ and ``confounder'' refers to variables $X_j$ with $\beta_{c,j} \neq 0$, $\beta_{d,j} \neq 0$. Here we refer to a confounder as any variable with $\beta_{c,j} \neq 0$, with the understanding that this is a {\em necessary} but not sufficient condition to be a confounding variable in the usual sense. Likewise, the term ``direct effect'' has other meanings in some related literature; here we will use it simply to refer to variables with $\beta_{d, j} \neq 0$. Moreover, our parametrization makes transparent how the linear regression (the response equation) ``controls for" confounding: the parameter $\alpha$ gives the rate of change in the response as a function of changes in treatment level due to ``random fluctuation'' ($\epsilon = Z  - \X\beta_c$). Intuitively, with $\beta_c$ and $\X$ in hand, we have access to a randomized experiment from which to infer $\alpha$. Crucially, the $Z$ likelihood enforces this interpretation of $\beta_c$ and hence $\alpha$.

Note also that this transformation leaves the likelihood unchanged. In particular, if one fits the selection equation via OLS and then substitutes the associated residuals into then response equation and then fits OLS, the resulting estimate of $\alpha$ will be exactly as if one used the original parameterization and fit the model via a single application of OLS. However, in terms of imposing regularization, the two parametrizations are quite different --- under our transformation the selection equation likelihood plays a role in dictating the degree of posterior shrinkage, because $\beta_c$ appears in both likelihoods.

Finally, note that this parametrization greatly mitigates the bias of $\hat{\alpha}$: given $\beta_c$, the expression for the bias under a flat prior for $\alpha$ and a standard normal prior for $\beta_d$ is
\begin{equation}\label{bias2}
\mbox{bias}(\hat{\alpha}) = -\left((\mathrm{r}^t\mathrm{r})^{-1}\mathrm{r}^t\mathbf{X}\right)(\mathbf{I}_p + \mathbf{X}^t(\mathbf{X} - \hat{\mathbf{X}}_R))^{-1}\beta_d.
\end{equation}
where $\mathrm{r} = \mathrm{z}-\mathbf{X}\beta_c$. By construction, $(\mathrm{R}^t\mathrm{R})^{-1}\mathrm{R}^t\mathbf{X}$ will be close to the zero vector, because $R_i = Z_i - X_i\beta_c$ is independent of $X_i$. Of course, $\beta_c$ (and hence $R$) is not known, but the new model is conditionally approximately unbiased for $\alpha$ and the $Z$ likelihood provides information on $\beta_c$. In fact, expressions (\ref{bias}) and (\ref{bias2}) indicate that the naive model will have higher bias the stronger the confounding (as measured by small $\sigma_{\epsilon}$), which is exactly when the new parametrization has more information about $
\beta_c$ and so will be closer to unbiased. This observation is borne out in the simulation studies below.

\section{Simulation studies}\label{simulations}
This section reports simulation studies which demonstrate the success of the reparametrized model in avoiding the mis-identified shrinkage of naive regularization. The four methods being compared are ordinary least squares (OLS) applied to the response equation, ``naive regularization" which applies a shrinkage prior over $\beta$ and uses only the response equation likelihood, the new approach, which places independent shrinkage priors over $\beta_c$ and $\beta_d$ and uses both the response and treatment likelihoods, and ``oracle OLS" which performs OLS using only the variables with non-zero coefficients. Note that oracle OLS is not possible to implement in applied problems. Non-informative priors over the remaining parameters are the same for both Bayesian approaches: $\alpha \propto 1$, $\sigma_{\epsilon} \propto 1/\sigma_{\epsilon}$,  $\sigma_{\nu} \propto 1/\sigma_{\nu}$.

In this paper, the shrinkage prior we employ is 
\begin{equation}\label{horseshoe}
\begin{split}
\pi(\beta_j) &\propto \frac{1}{v}\log{\left(1 + \frac{4}{(\beta_j/v)^2}\right)},\\
\pi(v) &\sim \mbox{C}^{+}(0,1),
\end{split}
\end{equation}
where $v$ is a global scale parameter common across all elements $j = 1, \dots p$, and $\mbox{C}^{+}(0,1)$ denotes a folded standard Cauchy distribution. This prior is a close proxy of the horseshoe prior of \cite{carvalho2010horseshoe}. Such priors have proven empirically to be a fine default choice for regression coefficients:  they lack hyperparameters, forcefully separate strong from weak predictors, and exhibit robust predictive performance. This modified representation permits the model to be fit using an elliptical slice sampler of \cite{hhl}; as reported there, when $p = 1000$ this sampler can produce 10,000 posterior samples in less than a minute (for any sample size strictly larger than $p$). We defer the computational details of our approach to the appendix.  We stress, however, that the key patterns revealed in our simulation study are a byproduct primarily of our reparameterization, and can be expected to arise under any similar regularization prior. Although not reported here, simulation studies were also conducted under ridge priors (with empirical Bayes selection of the shrinkage parameter) and the basic conclusions do not change under these variations. We also include one simulation study using point-mass model selection priors (using within-model $g$-priors) for applications where $p > n$.
 
\subsection{\cite{wang2012bayesian} simulations}

In this section, we consider two simulations from the analysis of \cite{wang2012bayesian}.  In the first simulation, the true model for the data is: $Y_{i} = \alpha Z_{i} + \beta_{1}X_{1i} + \beta_{2}X_{2i} + \epsilon_{i}$, where $i = 1, ..., 1000$ and $\epsilon_{i} \sim N(0,1)$.  The vector of treatment and covariates is distributed as $(Z_{i},X_{1i},X_{2i}) \sim N(0,\Sigma)$ where $\Sigma_{kk} = 1$ for $k=1,2,3$, $\Sigma_{12} = \Sigma_{21} = \rho$, and $\Sigma_{13}=\Sigma_{31}=\Sigma_{23}=\Sigma_{32}=0$.  The potential confounders are $(X_{1}, X_{2})$ with 49 additional independent random variables drawn from a standard normal. We set the parameters as $\rho=0.7$ and $\alpha=\beta_{1}=\beta_{2}=0.1$ and generate 1000 data sets for analysis.

The results from the first simulation are displayed in table \ref{domsim1}.  We show average bias, interval length, and mean squared error across all generated data sets as well as the probability of covering the true treatment effect (coverage).  

In the second simulation, a larger set of potential confounders is considered, and they are correlated with both the treatment and response variables.  The true model is: $Y_{i} = \alpha Z_{i} + \beta_{1}X_{1i} + \cdots + \beta_{14}X_{14i} + \epsilon_{i}$, where $i = 1, ..., 1000$ and $\epsilon_{i} \sim N(0,1)$.  The vector of treatment and covariates is distributed as $(Z_{i},X_{1i}, ...,X_{7i}) \sim N(0,\Sigma)$.  The covariance matrix $\Sigma$ is designed so that weak and strong correlations among the treatment and confounders exist.  We set $\Sigma_{kl} = 1$ if $k=l$ and $\Sigma_{kl}=\rho^{k+l-2}$ if $k \neq l$ and $k,l \in \{1,...,8\}$.  The remaining covariates are $(X_{8},..., X_{14})$ are drawn from a standard normal. The entire set of potential confounders is $\X_{1},...X_{14}$ with 43 additional random variables drawn from a standard normal.  Similar to the first simulation, we set the parameters as $\rho=0.7$ and $\alpha=\beta_{1}=\cdots= \beta_{14}=0.1$ and again consider 1000 replications of this data set.

In both simulations, naive regularization performs poorly in coverage and is severely biased.  The new approach successfully reduces bias and has comparable performance to OLS in coverage, statistical power as measured by interval length, and mean squared error. The similar performance of our method and OLS in this case is due to the relatively large sample size for the given signal-to-noise level (\cite{wang2012bayesian} report nearly identical results as well). In the following section, we construct a simulation that shows when the data generating process has certain realistic properties, the new approach can outperform OLS in interval length and mean squared error (while naive regularization continues to underperform).

\begin{table}[H]
\begin{center}
\begin{tabular}{|l|c|c|c|c|}
\hline
&Bias & Coverage&I.L.& MSE\\
\hline
New Approach  &  0.0024  &  0.959  &  0.1754  &  0.002 \\
OLS  &  0.0014  &  0.96  &  0.1786  &  0.002 \\
Naive Regularization  &  0.0479  &  0.35  &  0.0774  &  0.0053 \\
Oracle OLS  &  0.0015  &  0.958  &  0.1738  &  0.0019 \\
\hline
\end{tabular}
\end{center}
\caption{\cite{wang2012bayesian}: Simulation Study 1.}
\label{domsim1}
\end{table}

\begin{table}[H]
\begin{center}
\begin{tabular}{|l|c|c|c|c|}
\hline
&Bias & Coverage&I.L.& MSE\\
\hline
New Approach  &  0.0034  &  0.955  &  0.201  &  0.0027 \\
OLS  &  -0.002  &  0.956  &  0.2022  &  0.0026 \\
Naive Regularization  &  0.0822  &  0.597  &  0.1889  &  0.0097 \\
Oracle OLS  &  -1e-04  &  0.94  &  0.1985  &  0.0028 \\
\hline
\end{tabular}
\end{center}
\caption{\cite{wang2012bayesian}: Simulation Study 2.}
\label{domsim2}
\end{table}

\subsection{Further simulations: shrinkage estimation in the presence of confounding}

In this section, we show results from a simulation designed to capture a variety of scenarios a data analyst may face. We consider changing the relative strengths of the confounding and direct effects as well as the number of such variables.  Specifically, we use the two equation model (\ref{themodel}) to generate our data. We set the marginal variance of the treatment and response variables to one, $\mbox{var}(Z)= \mbox{var}(Y)=1$, and we center and scale the control variables $\X$ to have mean zero and unit variance.  

To ensure we consider a range of data compositions, we parametrize our simulations using an ANOVA style decomposition. Defining the $\ell$-2 norms (squared Euclidean distance) of the confounding and direct effects as $\rho^{2}=\| \beta_c \|_{2}^{2}$ and $\phi^{2}=\| \beta_d \|_{2}^{2}$, we may decompose the marginal variances as    
\begin{equation}
	\begin{split}\label{vardec}
		\mbox{var}(Z) &= \rho^{2} + \sigma_{\epsilon}^{2}
		\\
		\mbox{var}(Y) &= \alpha^{2}(1-\rho^{2}) + \phi^{2} + \sigma_{\nu}^{2},\\
		& = \kappa^2 + \phi^2 + \sigma_{\nu}^2,
	\end{split}
\end{equation}
because the control variables are standardized. Fixing the marginal variances to one implies $\sigma_{\epsilon}^{2} = 1 - \rho^{2}$ and $\sigma_{\nu}^{2} = 1 - \alpha^{2}(1-\rho^{2}) - \phi^{2}$. This decomposition admits the following interpretation: $\rho^{2}$ is the percentage of the treatment's variance due to confounding (strength of the confounding effect), $\phi^{2}$ is the percentage of the response variance due to the direct impact of the control variables on the response (strength of the direct effect), and $\kappa^2 := \alpha^{2}(1-\rho^{2})$ is the percentage of the response variance due to quasi-experimental variation of the treatment variable. 

Next, observe that as the confounding becomes stronger ($\rho^2$ getting larger), the independent variation from which we infer the treatment effect ($Z - \X\beta_c$) becomes smaller ($1 - \rho^2$). This means that for a fixed level of treatment effect, $\alpha$, and a fixed marginal variance, stronger confounding makes treatment effect inference harder in that the residual variance becomes correspondingly larger:  $1 - \alpha^2(1 - \rho^2) - \phi^2$. This makes it more difficult to get a clear picture of whether or not the confounding {\em per se} is making the problem difficult, or if problems with strong confounding just happen to be more difficult in this artificial way. To avoid this problem, we fix $\kappa^2 := \alpha^{2}(1-\rho^{2})$ to a constant, and allow $\alpha$ to vary as $\rho^2$ is varied. In this way we can examine the impact of confounding for a fixed difficulty of inference (as measured by the residual variance, which is held fixed at $1 - \kappa^2 - \phi^2$).

In our simulations, we fix a decomposition of the response variance given in \ref{vardec} and vary the strength of the confounding effect, $\rho^{2}$.  This amounts to specifying values for $\kappa^{2}$, $\phi^{2}$, and $\sigma_{\nu}^{2}$ that sum to one, and simulating data sets for several values  of $\rho^{2}$ between $0$ and $1$.  Again, because $\kappa^{2} = \alpha^{2}(1-\rho^{2})$ is fixed, as $\rho^{2}$ varies, $\alpha$ will vary as well. 

Next, the components of $\beta_c$ and $\beta_d$ must be specified.  The nonzero entries of each identify which $\X_{i}$'s are confounders, direct effects, and both, as previously defined. We define the first $k$ elements of $\X$ to be confounders, the next $k$ to be both confounders \textit{and} direct effects, and the final $k$ elements to be direct effects.  We achieve this in our simulation by setting $\beta_{c}^{1:2k}$ to ones and $\beta_{d}^{(k+1):3k} \sim \mbox{N}(0,1)$.  These vectors are then rescaled to have magnitudes $\rho^{2}$ and $\phi^{2}$, respectively.  This sets the overall $\beta$ vector ($\beta = \beta_d - \alpha \beta_c$) to have $3k$ nonzero entries. (Note that under continuous priors for $\beta_c$ and $\beta_d$, every variable is a confounder and no variables are strictly prognostic.) 

Let $n$ be the number of observations and $p$ be the number of columns of $\X$.  In our simulation, we set $n=100,50$ and $p=30$.  Additionally, we consider the following response variance decompositions: $\{\kappa^{2}=0.05, \phi^{2}=0.7, \sigma_{\nu}^{2}=0.25\}$, $\{\kappa^{2}=0.05, \phi^{2}=0.05, \sigma_{\nu}^{2}=0.9\}$ and vary $\rho^{2} \in \{0.1,0.3,0.5,0.7,0.9\}$.

Tables \ref{simtab1} and \ref{simtab2} show results for the variance decomposition $\{\kappa^{2}=0.05, \phi^{2}=0.7, \sigma_{\nu}^{2}=0.25\}$ and $n=100$ and $50$, respectively.  In this scenario, the direct effect drives 70\% of variance in the response while the treatment effect drives 5\%.  Tables displaying the numbers used to generate these plots are shown in the appendix.  Similar to the \cite{wang2012bayesian} example, we compare the new, OLS, and naive regularization approaches in the presence of weak to strong confounding ($\rho^{2} \in \{0.1,0.3,0.5,0.7,0.9\}$). Again, the oracle OLS result is given for comparison. The four metrics we evaluate are bias, mean squared error (MSE), interval length (I.L.), and coverage.  First, note the poor performance of the naive approach.  As confounding strength increases, bias grows and coverage decays exponentially for both sample sizes.  In addition, MSE explodes for increasing confounding strength.  Nevertheless, the naive approach does produce a small interval length resulting from the regularization prior.


As table \ref{simtab1} demonstrates, the new approach and OLS are comparable when the data size is large relative to the number of potential confounders with MSE and I.L. gains using the new approach when confounding strength is large ($\rho^{2}>0.9$).  When the data size is smaller (table \ref{simtab2}), the gains of using the new approach over OLS are seen across the board.   The new approach outperforms OLS in both interval length and MSE for confounding levels varying from weak to strong. This is the benefit of ``betting on sparsity" when the data generating process is in fact sparse.

Tables \ref{simtab3} and \ref{simtab4} show results for a different response variance configuration: $\{\kappa^{2}=0.05, \phi^{2}=0.05, \sigma_{\nu}^{2}=0.9\}$.  In this scenario, the treatment and direct effects contribute 5\% each to the response variance and the remaining 90\% is residual noise.  This is a problem that, using any method for estimation, is inferentially difficult because of the low signal-to-noise ratio of the response.  In both the large data set ($n=100$, table \ref{simtab3}) and small data set ($n=50$, table \ref{simtab4}) relative to the number of potential controls, we again see underperformance of the naive approach.

In contrast to the previous example with a strong direct effect, the weak direct effect contributes to good performance of the new approach relative to OLS for both $n=100$ and $n=50$. Again, we see that the new approach has increased power through smaller interval lengths and lower mean squared error, especially for data sets with strong confounding. And again, we see the benefit of ``betting on sparsity" when the data generating process is in fact sparse.

\begin{table}[H]
\begin{center}
\begin{tabular}{|l|l|c|c|c|c|}
\hline
$\rho^{2}$ & & Bias & Coverage & I.L. & MSE \\
\hline 
0.1  &  New Approach  &  -0.0032  &  0.943  &  0.2357  &  0.0037 \\
 &  OLS  &  -0.0016  &  0.951  &  0.2477  &  0.004 \\
 &  Naive Regularization  &  -0.0112  &  0.895  &  0.2089  &  0.0037 \\
 &  Oracle OLS  &  0.0023  &  0.946  &  0.2173  &  0.0031 \\

\hline 
0.3  &  New Approach  &  -0.0047  &  0.95  &  0.2751  &  0.0047 \\
 &  OLS  &  -0.0018  &  0.951  &  0.2808  &  0.0052 \\
 &  Naive Regularization  &  -0.0355  &  0.848  &  0.2293  &  0.0057 \\
 &  Oracle OLS  &  0.0026  &  0.946  &  0.2464  &  0.004 \\

\hline 
0.5  &  New Approach  &  -3e-04  &  0.963  &  0.3345  &  0.0066 \\
 &  OLS  &  -0.0022  &  0.951  &  0.3323  &  0.0072 \\
 &  Naive Regularization  &  -0.0768  &  0.746  &  0.2631  &  0.012 \\
 &  Oracle OLS  &  0.0031  &  0.946  &  0.2915  &  0.0056 \\

\hline 
0.7  &  New Approach  &  0.0084  &  0.964  &  0.4374  &  0.0113 \\
 &  OLS  &  0.0024  &  0.944  &  0.4303  &  0.0123 \\
 &  Naive Regularization  &  -0.1559  &  0.543  &  0.3292  &  0.0346 \\
 &  Oracle OLS  &  0.004  &  0.946  &  0.3764  &  0.0093 \\

\hline 
0.9  &  New Approach  &  -0.004  &  0.972  &  0.7403  &  0.0292 \\
 &  OLS  &  0.0045  &  0.954  &  0.7469  &  0.0351 \\
 &  Naive Regularization  &  -0.4482  &  0.231  &  0.4779  &  0.2391 \\
 &  Oracle OLS  &  0.0069  &  0.946  &  0.6519  &  0.0278 \\
\hline
\end{tabular}
\end{center}
\caption{$\mathbf{n=100,p=30,k=3}$. $\kappa^{2}=0.05$. $\phi^{2}=0.7$. $\sigma_{\nu}^{2}=0.25$.}
\label{simtab1}
\end{table}

\begin{table}[H]
\begin{center}
\begin{tabular}{|l|l|c|c|c|c|}
\hline
$\rho^{2}$ & & Bias & Coverage & I.L. & MSE \\
\hline 
0.1  &  New Approach  &  0.0082  &  0.918  &  0.3632  &  0.0105 \\
 &  OLS  &  -0.0017  &  0.944  &  0.4785  &  0.0144 \\
 &  Naive Regularization  &  -0.0068  &  0.835  &  0.2957  &  0.0097 \\
 &  Oracle OLS  &  -0.001  &  0.952  &  0.3235  &  0.0065 \\

\hline 
0.3  &  New Approach  &  -1e-04  &  0.94  &  0.4203  &  0.0128 \\
 &  OLS  &  -0.002  &  0.944  &  0.5425  &  0.0186 \\
 &  Naive Regularization  &  -0.035  &  0.837  &  0.3191  &  0.0126 \\
 &  Oracle OLS  &  -0.0011  &  0.952  &  0.3668  &  0.0084 \\

\hline 
0.5  &  New Approach  &  -0.0047  &  0.93  &  0.5183  &  0.0196 \\
 &  OLS  &  -0.0023  &  0.944  &  0.6419  &  0.026 \\
 &  Naive Regularization  &  -0.0869  &  0.738  &  0.3555  &  0.0222 \\
 &  Oracle OLS  &  -0.0014  &  0.952  &  0.434  &  0.0117 \\

\hline 
0.7  &  New Approach  &  0.0056  &  0.937  &  0.6926  &  0.0341 \\
 &  OLS  &  0.0046  &  0.934  &  0.8204  &  0.0478 \\
 &  Naive Regularization  &  -0.189  &  0.539  &  0.4033  &  0.0565 \\
 &  Oracle OLS  &  -0.0018  &  0.952  &  0.5604  &  0.0195 \\

\hline 
0.9  &  New Approach  &  -0.0772  &  0.959  &  1.1572  &  0.0804 \\
 &  OLS  &  -0.0156  &  0.931  &  1.4347  &  0.1402 \\
 &  Naive Regularization  &  -0.5419  &  0.102  &  0.4868  &  0.3297 \\
 &  Oracle OLS  &  -0.003  &  0.952  &  0.9706  &  0.0585 \\
\hline
\end{tabular}
\end{center}
\caption{$\mathbf{n=50,p=30,k=3}$. $\kappa^{2}=0.05$. $\phi^{2}=0.7$. $\sigma_{\nu}^{2}=0.25$.}
\label{simtab2}
\end{table}

\begin{table}[H]
\begin{center}
\begin{tabular}{|l|l|c|c|c|c|}
\hline
$\rho^{2}$ & & Bias & Coverage & I.L. & MSE \\
\hline 
0.1  &  New Approach  &  -0.0053  &  0.93  &  0.4137  &  0.0126 \\
 &  OLS  &  -0.0031  &  0.951  &  0.4699  &  0.0145 \\
 &  Naive Regularization  &  -0.027  &  0.434  &  0.1528  &  0.0178 \\
 &  Oracle OLS  &  0.0044  &  0.946  &  0.4123  &  0.0111 \\

\hline 
0.3  &  New Approach  &  -0.0101  &  0.933  &  0.472  &  0.0176 \\
 &  OLS  &  -0.0035  &  0.951  &  0.5329  &  0.0186 \\
 &  Naive Regularization  &  -0.075  &  0.373  &  0.1625  &  0.0256 \\
 &  Oracle OLS  &  0.005  &  0.946  &  0.4675  &  0.0143 \\

\hline 
0.5  &  New Approach  &  -0.001  &  0.933  &  0.5751  &  0.0245 \\
 &  OLS  &  -0.0041  &  0.951  &  0.6305  &  0.026 \\
 &  Naive Regularization  &  -0.1407  &  0.304  &  0.1751  &  0.0411 \\
 &  Oracle OLS  &  0.0059  &  0.946  &  0.5532  &  0.02 \\

\hline 
0.7  &  New Approach  &  0.0044  &  0.95  &  0.7509  &  0.0368 \\
 &  OLS  &  -0.0049  &  0.953  &  0.8156  &  0.0394 \\
 &  Naive Regularization  &  -0.265  &  0.134  &  0.1801  &  0.0918 \\
 &  Oracle OLS  &  0.0076  &  0.946  &  0.7141  &  0.0333 \\

\hline 
0.9  &  New Approach  &  -0.01  &  0.939  &  1.2784  &  0.1131 \\
 &  OLS  &  -0.0022  &  0.942  &  1.416  &  0.1345 \\
 &  Naive Regularization  &  -0.6114  &  0.002  &  0.1841  &  0.3983 \\
 &  Oracle OLS  &  0.0132  &  0.946  &  1.2369  &  0.0999 \\
\hline
\end{tabular}
\end{center}
\caption{$\mathbf{n=100,p=30,k=3}$. $\kappa^{2}=0.05$. $\phi^{2}=0.05$. $\sigma_{\nu}^{2}=0.9$.}
\label{simtab3}
\end{table}

\begin{table}[H]
\begin{center}
\begin{tabular}{|l|l|c|c|c|c|}
\hline
$\rho^{2}$ & & Bias & Coverage & I.L. & MSE \\
\hline 
0.1  &  New Approach  &  0.0021  &  0.919  &  0.6073  &  0.0306 \\
 &  OLS  &  -0.0119  &  0.93  &  0.8888  &  0.0528 \\
 &  Naive Regularization  &  -0.0291  &  0.443  &  0.2207  &  0.0352 \\
 &  Oracle OLS  &  -0.0019  &  0.952  &  0.6138  &  0.0234 \\

\hline 
0.3  &  New Approach  &  -0.0033  &  0.909  &  0.6918  &  0.0421 \\
 &  OLS  &  -0.0038  &  0.944  &  1.0294  &  0.0668 \\
 &  Naive Regularization  &  -0.0651  &  0.402  &  0.237  &  0.0428 \\
 &  Oracle OLS  &  -0.0022  &  0.952  &  0.696  &  0.0301 \\

\hline 
0.5  &  New Approach  &  -0.011  &  0.894  &  0.8191  &  0.064 \\
 &  OLS  &  0.0071  &  0.927  &  1.2041  &  0.103 \\
 &  Naive Regularization  &  -0.1354  &  0.349  &  0.233  &  0.0577 \\
 &  Oracle OLS  &  -0.0026  &  0.952  &  0.8235  &  0.0421 \\

\hline 
0.7  &  New Approach  &  -0.028  &  0.904  &  1.0842  &  0.105 \\
 &  OLS  &  -8e-04  &  0.938  &  1.5533  &  0.1603 \\
 &  Naive Regularization  &  -0.2752  &  0.217  &  0.2474  &  0.1163 \\
 &  Oracle OLS  &  -0.0033  &  0.952  &  1.0632  &  0.0702 \\

\hline 
0.9  &  New Approach  &  -0.1078  &  0.948  &  1.8128  &  0.2303 \\
 &  OLS  &  0.0291  &  0.942  &  2.6708  &  0.4893 \\
 &  Naive Regularization  &  -0.6045  &  0.015  &  0.2576  &  0.4096 \\
 &  Oracle OLS  &  -0.0058  &  0.952  &  1.8415  &  0.2106 \\
\hline
\end{tabular}
\end{center}
\caption{$\mathbf{n=50,p=30,k=3}$. $\kappa^{2}=0.05$. $\phi^{2}=0.05$. $\sigma_{\nu}^{2}=0.9$.}
\label{simtab4}
\end{table}

\subsection{Dense case}
In tables \ref{densetable1} and \ref{densetable2}, the same simulation study as before is run with $\{p = 30,\hspace{1mm} k = 10\}$ and $\{p = 30, \hspace{1mm} k = 30\}$, respectively.  For the $k = 10$ case, $\beta_c$ and $\beta_d$ each have 20 nonzero entries and 10 zero entries and are sparse with respect to our transformed model \ref{themodel}.  However, $\beta$ itself is dense. For the $k=30$ case, we abuse our simulation construction slightly and construct both $\beta_c$ and $\beta_d$ (and thus $\beta$) as fully dense vectors with all $p=30$ components nonzero.  In both cases, note that OLS and Oracle OLS are identical methods. Two salient patterns emerge from this simulation. First, the new method performs essentially on par with OLS; there is no benefit for the bet on sparsity, but their is no major penalty either. Second, the naive response-only regularized regression continues to exhibit dismal performance.

\begin{table}[H]
\begin{center}
\begin{tabular}{|l|l|c|c|c|c|}
\hline
$\rho^{2}$ & & Bias & Coverage & I.L. & MSE \\
\hline 
0.1  &  New Approach  &  -0.0038  &  0.939  &  0.2484  &  0.0043 \\
 &  OLS  &  -0.0014  &  0.944  &  0.2497  &  0.0041 \\
 &  Naive Regularization  &  -0.0094  &  0.948  &  0.241  &  0.0039 \\
 &  Oracle OLS  &  -0.0014  &  0.944  &  0.2497  &  0.0041 \\

\hline 
0.3  &  New Approach  &  -0.0051  &  0.94  &  0.2895  &  0.0057 \\
 &  OLS  &  0.0029  &  0.929  &  0.2827  &  0.0057 \\
 &  Naive Regularization  &  -0.0268  &  0.921  &  0.2638  &  0.0055 \\
 &  Oracle OLS  &  0.0029  &  0.929  &  0.2827  &  0.0057 \\

\hline 
0.5  &  New Approach  &  -0.012  &  0.966  &  0.351  &  0.007 \\
 &  OLS  &  -0.001  &  0.946  &  0.3327  &  0.007 \\
 &  Naive Regularization  &  -0.0715  &  0.85  &  0.2964  &  0.0103 \\
 &  Oracle OLS  &  -0.001  &  0.946  &  0.3327  &  0.007 \\

\hline 
0.7  &  New Approach  &  -0.0105  &  0.96  &  0.4614  &  0.0126 \\
 &  OLS  &  -1e-04  &  0.946  &  0.4279  &  0.0124 \\
 &  Naive Regularization  &  -0.1587  &  0.563  &  0.3489  &  0.0341 \\
 &  Oracle OLS  &  -1e-04  &  0.946  &  0.4279  &  0.0124 \\

\hline 
0.9  &  New Approach  &  -0.0496  &  0.963  &  0.7862  &  0.0351 \\
 &  OLS  &  -0.012  &  0.953  &  0.748  &  0.0369 \\
 &  Naive Regularization  &  -0.5131  &  0.01  &  0.4303  &  0.2764 \\
 &  Oracle OLS  &  -0.012  &  0.953  &  0.748  &  0.0369 \\  
 \hline
 \end{tabular}
\end{center}
\caption{$\mathbf{n=100,p=30,k=10}$. $\kappa^{2}=0.05$. $\phi^{2}=0.7$. $\sigma_{\nu}^{2}=0.25$.}
\label{densetable1}
\end{table}

\begin{table}[H]
\begin{center}
\begin{tabular}{|l|l|c|c|c|c|}
\hline
$\rho^{2}$ & & Bias & Coverage & I.L. & MSE \\
\hline 
0.1  &  New Approach  &  0.0023  &  0.942  &  0.2563  &  0.0045 \\
 &  OLS  &  0.0013  &  0.945  &  0.249  &  0.0041 \\
 &  Naive Regularization  &  -0.0025  &  0.947  &  0.2525  &  0.004 \\
 &  Oracle OLS  &  0.0013  &  0.945  &  0.249  &  0.0041 \\

\hline 
0.3  &  New Approach  &  -0.0048  &  0.954  &  0.2996  &  0.0057 \\
 &  OLS  &  -0.0039  &  0.956  &  0.2841  &  0.0052 \\
 &  Naive Regularization  &  -0.0215  &  0.937  &  0.28  &  0.0056 \\
 &  Oracle OLS  &  -0.0039  &  0.956  &  0.2841  &  0.0052 \\

\hline 
0.5  &  New Approach  &  0.003  &  0.965  &  0.3653  &  0.0078 \\
 &  OLS  &  9e-04  &  0.952  &  0.3334  &  0.0071 \\
 &  Naive Regularization  &  -0.0411  &  0.905  &  0.3171  &  0.0091 \\
 &  Oracle OLS  &  9e-04  &  0.952  &  0.3334  &  0.0071 \\

\hline 
0.7  &  New Approach  &  0.0042  &  0.954  &  0.4813  &  0.0149 \\
 &  OLS  &  3e-04  &  0.929  &  0.4328  &  0.014 \\
 &  Naive Regularization  &  -0.1147  &  0.772  &  0.3854  &  0.025 \\
 &  Oracle OLS  &  3e-04  &  0.929  &  0.4328  &  0.014 \\

\hline 
0.9  &  New Approach  &  -0.0329  &  0.965  &  0.8018  &  0.0363 \\
 &  OLS  &  -0.0053  &  0.942  &  0.7433  &  0.0375 \\
 &  Naive Regularization  &  -0.4212  &  0.155  &  0.5178  &  0.2052 \\
 &  Oracle OLS  &  -0.0053  &  0.942  &  0.7433  &  0.0375 \\  
 \hline
 \end{tabular}
\end{center}
\caption{$\mathbf{n=100,p=30,k=30}$. $\kappa^{2}=0.05$. $\phi^{2}=0.7$. $\sigma_{\nu}^{2}=0.25$.}
\label{densetable2}
\end{table}

\subsection{$p>n$ case}

In order to explore the behavior of our proposal in a $p>n$ set-up,  we extend the first simulation analysis of \cite{wang2012bayesian}.  Now, the true model for the data is: $Y_{i} = \alpha Z_{i} + \beta_{1}X_{1i} + \beta_{2}X_{2i} + \epsilon_{i}$, where $i = 1, ..., 30$ and $\epsilon_{i} \sim N(0,0.04)$.  The vector of treatment and covariates is distributed as $(Z_{i},X_{1i},X_{2i}) \sim N(0,\Sigma)$ where $\Sigma_{kk} = 1$ for $k=1,2,3$, $\Sigma_{12} = \Sigma_{21} = \rho$, and $\Sigma_{13}=\Sigma_{31}=\Sigma_{23}=\Sigma_{32}=0$.  The potential confounders are $(X_{1}, X_{2})$ with 33 additional independent random variables drawn from a standard normal, for a total of 35 control variables. We set the parameters as $\rho=0.7$ and $\alpha=\beta_{1}=\beta_{2}=0.1$ and generate 1000 data sets for analysis.

In the $p>n$ setting it is helpful to return to a variable-selection model. Specifically, we employ normal $g$-priors \citep{zellner1986assessing} on both $\beta_c$ and $\beta_d$ with point-masses at zero. We define $g$ through {\em local empirical Bayes} \citep{liang2012mixtures} with model probabilities are defined by $p(\mathcal{M})\propto {{p}\choose{2}} ^{-1} \mathbf{1}_{p< p_{\mbox{max}}}$ where $p_{\mbox{max}}$ defines the maximum number of non-zero elements in both $\beta_c$ and $\beta_d$ (separately). 

The primary reason for adopting this model in this setting is that it allows to directly handle exact-sparsity via the $p_{\mbox{max}}$ parameter; we can examine how the method behaves as this parameter changes relative to the true level of sparsity (two non-zero elements out of 35, in this case). A secondary reason is that the elliptical slice sampler we use for the continuous prior model would require special modification for the $p > n$ setting, because the maximum likelihood estimate is not well-defined. As a side benefit, this simulation allows us to demonstrate and emphasize that the benefits of the new parameterization are fundamentally prior-agnostic; it is not the specific choice of prior that matter, rather it is the ability to specify the prior in terms of $\beta_c$ and $\beta_d$.  It is worth noting that the local empirical Bayes approach can be quite slow when $p_{\mbox{max}}$ is large; when $p_{\mbox{max}} = 1000$ it will take more than a dozen minutes to obtain ten thousand samples (whereas the horseshoe implementation discussed above would take approximately one minute).

Table \ref{pgreater}  shows the results of this study, where the same prior is used for $\beta$ (the naive approach) versus separately for both $\beta_c$ and $\beta_d$ in the new parametrization. The new model performs best when $p_{\mbox{max}}$ is smaller (closer to the true number of non-zero coefficients), according to mean squared error. In all cases except $p_{\mbox{max}} = 20$, the MSE of the new model is lower than the naive model. In every case, the new model has better coverage. The naive model has smaller posterior credible intervals, but greater bias.

\begin{table}[H]
\begin{center}
\begin{tabular}{|l|c|c|c|c|}
\hline
$p=35$, $n=30$. $p_{\mbox{max}}=3$&Bias & Coverage&I.L.& MSE\\
\hline
New Approach  &  0.055  &  0.87 &  0.301  &  0.008 \\
Naive Regularization  &  0.093  &  0.64  &  0.239  &  0.012 \\
\hline
$p=35$, $n=30$. $p_{\mbox{max}}=5$&Bias & Coverage&I.L.& MSE\\
\hline
New Approach  &  0.056  &  0.88  &  0.319  &  0.010 \\
Naive Regularization  &  0.097  &  0.60  &  0.239  &  0.013 \\
\hline
$p=35$, $n=30$. $p_{\mbox{max}}=10$&Bias & Coverage&I.L.& MSE\\
\hline
New Approach  &  0.059 &  0.88  &  0.335  &  0.010 \\
Naive Regularization  &  0.099  &  0.63  &  0.255  &  0.013 \\
\hline
$p=35$, $n=30$. $p_{\mbox{max}}=20$&Bias & Coverage&I.L.& MSE\\
\hline
New Approach  &  0.068 &  0.86  &  0.435  &  0.016 \\
Naive Regularization  &  0.103  &  0.65  &  0.255  &  0.015 \\
\hline
\end{tabular}
\end{center}
\caption{A variable selection prior used in the $p > n$ setting still reveals the benefit of the new parameterization over the naive response-only model. In this simulation the true data generating process had only two non-zero regression coefficients; accordingly, the model performs better when $p_{\mbox{max}}$ is smaller, according to mean squared error (MSE). In all cases except $p_{\mbox{max}} = 20$, the MSE is lower than the naive model. In every case, the new model has better coverage. The naive model has smaller posterior credible intervals, but greater bias.}\label{pgreater}
\end{table}

\section{Empirical illustration: abortion and crime}\label{levitt}
In this section, we consider the relationship between legalized abortion and crime rates, using data first analyzed in \cite{donohue2001impact} and widely publicized in the popular book \cite{levitt2010freakonomics}.  \cite{donohue2001impact} propose that their data tell an intriguing story: unwanted children are more likely to grow up to be criminals, so legalized abortion, which leads to fewer unwanted children, leads to lower levels of crime in society. They conduct three analyses, one each for three different types of crime: violent crime, property crime, and murders.

Here, in the spirit of the similar reanalysis of \cite{belloni2014inference}, we reanalyze the \cite{donohue2001impact} data using a substantially more elaborate model, and observe the impact regularization has on the resulting conclusions. Specifically, we will compare four estimation approaches: one using the original control variables and OLS, one using an expanded covariate set (which includes many interactions) fit with OLS, one using the expanded covariate set fit with a naively regularized Bayesian regression, and one using the expanded covariate set fit with a regularized Bayesian model using our new parametrization. 

The response variable, $Y$, is per capita crime rates (violent crime, property crime, and murders) by state, from 1985 to 1997 (inclusive). The treatment variable, $Z$, is the ``effective'' abortion rate. This metric is an averaged abortion rate, weighted by criminal age at the time of arrest (to account for the fact that crimes committed by criminals should be associated with abortion rates at the time of their births). 

As control variables, $\X$, \cite{donohue2001impact} include a host of state and year specific attributes that could otherwise contribute to the observed crime rates:
\begin{itemize}
 \item prisoners per capita (log),
\item  police per capita (log),
\item state unemployment rate, 
\item state income per capita (log), 
\item percent of population below the poverty line, 
\item generosity of Aid to Families with Dependent Children (lagged by fifteen years), 
\item concealed weapons law, 
\item beer consumption per capita.
 \end{itemize}
Including state and year dummy variables brings the total number of control variables to 66. For additional details concerning how these attributes are defined and where they were obtained, see the original paper \citep{donohue2001impact}.

Our expanded model includes the following additional control variables:
\begin{itemize}
\item  interactions between the original eight controls and year, 
\item interactions between the original eight controls and year squared,
\item interactions between state effects and year,
\item interactions between state effects and year squared.
\end{itemize}

These additional variables allow the impact of the original eight covariates on crime rate to change flexibly across time (according to a quadratic trend) and allows for the state specific crime rates to likewise change over time (in terms of an offset from overall state and year rates according to a quadratic trend). When allowing for this degree of flexibility, estimation becomes quite challenging, with just $n = 624$ observations and $p = 176$ control variables.

Tables \ref{LevittTable1} and \ref{LevittTable2} show our posterior inference compared to OLS and naive regularization. First, we note that the reported OLS results on the original covariate set are very similar to the results given in \cite{donohue2001impact}, although they used weighted least squares to adjust for differing state populations. 

Second, using the original covariate set, the results of our new method are broadly in agreement with OLS. Already in this case we observe signs of the naive regularization approach being biased.

Finally, using the augmented covariate set, we observe that OLS no longer identifies the originally reported negative effect. However, the interval it returns is not tight about zero, indicating that there is not enough signal in the data to determine the impact of abortion on crime rates. Our new approach, by comparison, has much smaller credible interval, although they also include zero. Notably, the asymmetry (with respect to zero) of the interval of OLS and our approach coincide, while naive regularization is off-centered in the opposite direction. In fact, naive regularization excludes zero in the case of property and violent crime, and reports the reverse of the effect in the original study. This relationship between the three methods bears out the patterns observed in our simulation studies and suggests that the naively regularized method is misestimating the treatment effect as a result of misallocated shrinkage.

\begin{table}
\begin{tabular}{l|cc||cc||cc||}
\cline{2-7}
&\multicolumn{2}{c||}{Property Crime}&\multicolumn{2}{c||}{Violent Crime}&\multicolumn{2}{c||}{Murder}\\
\cline{2-7}
&2.5\% &97.5\%&2.5\% &97.5\%&2.5\% &97.5\%\\
\hline
OLS   &-0.110 & -0.072& -0.171 &-0.090& -0.221& -0.040\\ 
new approach &-0.113 & -0.073&-0.182&-0.098&-0.222 &-0.039  \\
naive regularization &-0.075 & -0.010&0.079& 0.301&-0.186 &0.085  \\
\hline
\end{tabular}
\caption{\label{LevittTable1}Credible/confidence intervals (95\%) for the Donohue and Levitt (2001) example with original controls ($p = 66$, $n = 624$). On the smaller set of original controls, our new approach gives similar credible intervals as the OLS confidence interval. In this case, already the naive regularization approach shows signs of bias, although the impact is minor.}
\end{table}

\begin{table}
\begin{tabular}{l|cc||cc||cc||}
\cline{2-7}
&\multicolumn{2}{c||}{Property Crime}&\multicolumn{2}{c||}{Violent Crime}&\multicolumn{2}{c||}{Murder}\\
\cline{2-7}
&2.5\% &97.5\%&2.5\% &97.5\%&2.5\% &97.5\%\\
\hline
OLS & -0.226 & 0.019  & -0.374 &0.336&-0.125 & 1.763 \\
new approach  &-0.038 & 0.014& -0.114 &0.053&-0.081 &0.279 \\
naive regularization  & 0.007 &0.129 & 0.011 &0.412 &-0.227 &0.116\\
\hline
\end{tabular}
\caption{\label{LevittTable2}Credible/confidence intervals (95\%) for the Donohue and Levitt (2001) example with augmented controls ($p = 176$, $n = 624$). With the enlarged set of control variables, the new approach and OLS show notable differences, specifically our new regularized Bayesian approach has markedly smaller credible intervals. The naive regularization approach disagrees on the directionality of the effect compared to the other two methods, consistent with the bias observed in our simulation studies.}
\end{table}

\section{Discussion}
In this paper, we have documented the perhaps counterintuitive fact that naively applied shrinkage priors can dramatically corrupt inference concerning treatment effects and have developed a regularized Bayesian regression model that avoids this pitfall, while still boasting the usual advantages of shrinkage estimation. 

In this section we conclude with additional discussion concerning the mechanism by which this parametrization improves estimation. Specifically, while it is explained above that the new parametrization is designed to be approximately unbiased for $\alpha$ (as a function of $\beta$), it is perhaps less clear that shrinkage priors on $\beta_c$ and $\beta_d$ are not conferring some additional advantage. For example, adjusting for variables that only associate with $Z$, but not with $Y$, is widely understood to decrease precision in estimates of $\alpha$ (relative to the model that omits these variables from the regression altogether). This phenomenon can be understood concretely through the lens of the new parametrization. First, such variables have a direct parametric interpretation: $\beta_d = 0$ and $\beta_c \neq 0$. Now, suppose that someone informs the analysts that a certain $\beta_d = 0$ {\em a priori}; in this event, one is better off running a regression to estimate $\alpha$ excluding the variable $X_j$ from the model because, intuitively, larger variation in the implied residuals gives more heterogeneity to estimate $\alpha$. 

In fact, this intuition can be made more precise. Without loss of generality consider the case of only one potential confounder, $X$. If $\beta_c$ and $\beta_d$ were both known, consider estimating $$\alpha = \E\left((Y - X\beta_d)(Z - X\beta_c)\right)$$ from a sample $(Y_i, Z_i, X_i)$, $i = 1...n$. Some straightforward manipulation shows that $$\alpha = \E(YZ) - E(X^2)\beta_c\beta_d.$$ In the Gaussian linear regression model, the sample moments $n^{-1}\sum_i Y_iZ_i$ and $n^{-1}\sum_i X_i^2$ are sufficient statistics. From the above expression we observe that knowing that $\beta_d = 0$ annihilates the second term involving data $n^{-1}\sum_i X_i^2$; the associated estimator has less sample variability because it is unaffected by sampling variation in $n^{-1}\sum_i X_i^2$. However, any model that must estimate $\beta_d$ necessarily incorporates $n^{-1}\sum_i X_i^2$ and pays the price in precision. Therefore, the more prior mass about $\beta_c\beta_d = 0$, the more limited will be the sampling variation due to $n^{-1}\sum_i X_i^2$; independent zero-centered priors over $\beta_c$ and $\beta_d$ achieve that, while the use of fat-tailed priors allows the data to speak.

At the same time, better estimate of $\beta_c$ is naturally obtained by incorporating the sampling model for $Z$. Indeed, consider the case where $\beta_d$ is known and non-zero; one need not use $Z$ to obtain a consistent estimate of $\beta_c$ and $\alpha$, but discarding the $Z$ model (presuming it is correctly specified) is simply throwing away available information, as $\beta_c$ appears there. This is true especially if the signal-to-noise ratio in the selection equation is much more favorable than that of the response equation ($\sigma_{\nu} \gg \sigma_{\epsilon}$). This is precisely why our new parametrization pays dividends, because the naive parametrization implies that the $Z$ model is ignored. In other words, the new parametrization has an advantage over single-equation approaches in terms of estimating $\beta_c$, but not in terms of estimating $\beta_d$; for which no essentially new data is being brought to bear, merely a strongly zero-biased prior. (The extent to which this zero-bias is beneficial will presumably depend on the true data generating process; this is the subject of ongoing investigation.)

Ongoing work looks at adapting the ideas in this paper to the nonlinear regression models for treatment effect estimation; preliminary results are promising.

\bibliographystyle{imsart-nameyear.bst} 
\bibliography{RegularizedTreatmentEffect}
\begin{appendix}
\section{Posterior sampling the regularized treatment effect linear model}
We first describe our sampler in terms of a standard Gaussian linear model
$$Y_i = \mbox{X}_i\beta + \epsilon_i, \;\;\; \epsilon_i \sim \N(0,\sigma^2)$$
with arbitrary prior $\pi(\beta)$. At one level, our approach is a Gibbs sampler, alternating between sampling $\beta$ and $\sigma^2$. We omit the update for $\sigma^2 \mid \beta$, noting simply that under the non-informative priors used in this paper, the update is standard conjugate inverse-gamma form. For the $\beta \mid \sigma^2$ update, we use an elliptical slice sampler, described here.

Let $\hat{\beta} = (\mX^t \mX)^{-1}\mX^t \Y$ denote the least squares solution and for an initial value of $\beta$, define $\Delta := \beta - \hat{\beta}.$ We then sample $\beta$ as a vector, according to the following algorithm.

\begin{enumerate}
\item Draw $\zeta \sim \N(0, \sigma^2 (\mX ^t\mX)^{-1})$.
\item For $\upsilon \sim \mbox{Uniform}(0,1)$ define $\ell := \log{(\pi(\beta))} + \log{(\upsilon)}$.
\item Draw angle $\varphi \sim \mbox{Uniform}(0, 2\pi)$; set $lower \leftarrow \varphi - 2\pi$ and $upper \leftarrow \varphi$.
\item Set $\Delta' \leftarrow \Delta \cos{\varphi} + \zeta \sin{\varphi}$ and $\beta' \leftarrow \hat{\beta} + \Delta'$ .
\item {\bf while} $\log{(\pi(\beta'))} < \ell$
\begin{enumerate}
\item {\bf if} $\varphi < 0$, set $lower \leftarrow \varphi$, {\bf else} set $upper \leftarrow \varphi$.
\item Draw angle $\varphi \sim \mbox{Uniform}(lower, upper)$
\item Update $\Delta' \leftarrow \Delta \cos{\varphi} + \zeta \sin{\varphi}$ and $\beta' \leftarrow \hat{\beta} + \Delta'$.
\end{enumerate}
\item Set $\Delta \leftarrow \Delta'$ and $\beta \leftarrow \hat{\beta} + \Delta'$.\\
\end{enumerate}


Observe that this sampler draws from the posterior for $\beta$ under a flat prior and then adjust these draws according to the prior distribution, rotating the vector along level curves of the posterior (equivalently, likelihood) until an improvement is reached with respect to the prior.

In this paper, we have $\pi(\beta \mid v) = \prod_{j =1}^p \log{(1 + 4/(\beta_j/v)^2)}/v$ and we sample the hyperparameter $v$ (the so-called ``global" scale parameter) within our Gibbs sampler as a random walk Metropolis update. 

To apply this sampler in the treatment effect context, we consider two cases. In the first case, the naive regularization approach, simply set $\mbox{D} = \mbox{X}_1$ and $\alpha = \beta_1$ and use the same prior but omitting $\beta_1$ from the prior evaluation, corresponding to $\pi(\alpha) \propto 1$.

The re-parametrized model proceeds analogously, but now we must jointly sample $(\alpha, \beta^t, \gamma^t)$ under flat priors and transform to $\beta_c$ and $\beta_d$ before prior evaluation. For initial values of the parameter, this gives:

\begin{enumerate}
\item Draw $\zeta_1 \sim \N(0, \sigma_{\nu}^2 (\mX^t \mX)^{-1})$ and  $\zeta_2 \sim \N(0, \sigma_{\epsilon}^2 (\mX^t \mX)^{-1})$ and defining $\zeta^t = (\zeta_1^t, \zeta_2^t)$. 
\item For $\upsilon \sim \mbox{Uniform}(0,1)$ define $\ell := \log{(\pi(\beta_c))} + \log{(\pi(\beta_d))} + \log{(\upsilon)}$.
\item Draw angle $\varphi \sim \mbox{Uniform}(0, 2\pi)$; set $lower \leftarrow \varphi - 2\pi$ and $upper \leftarrow \varphi$.
\item Set $\Delta' \leftarrow \Delta \cos{\varphi} + \zeta \sin{\varphi}$, $\alpha' \leftarrow \hat{\alpha} + \Delta_1'$, $\beta' \leftarrow \hat{\beta} + \Delta_{2:(p+1)}'$, $\gamma' \leftarrow \hat{\gamma} + \Delta_{(p+2):(2p + 1)}'$; with $\beta_c' = \gamma'$ and $\beta_d' = \alpha' \gamma' + \beta'$. 
\item {\bf while} $\log{(\pi(\beta_c'))} + \log{(\pi(\beta_d'))} < \ell$
\begin{enumerate}
\item {\bf if} $\varphi < 0$, set $lower \leftarrow \varphi$, {\bf else} set $upper \leftarrow \varphi$.
\item Draw angle $\varphi \sim \mbox{Uniform}(lower, upper)$
\item Update $\Delta' \leftarrow \Delta \cos{\varphi} + \zeta \sin{\varphi}$ and $\alpha' \leftarrow \hat{\alpha} + \Delta_1'$, $\beta' \leftarrow \hat{\beta} + \Delta_{2:(p+1)}'$, $\gamma' \leftarrow \hat{\gamma} + \Delta_{(p+2):(2p + 1)}'$; with $\beta_c' = \gamma'$ and $\beta_d' = \alpha' \gamma' + \beta'$.
\end{enumerate}
\item Set $\Delta \leftarrow \Delta'$ and $\alpha \leftarrow \hat{\alpha} + \Delta_1'$, $\beta \leftarrow \hat{\beta} + \Delta_{2:(p+1)}'$, $\gamma \leftarrow \hat{\gamma} + \Delta_{(p+2):(2p + 1)}'$; with $\beta_c = \gamma$ and $\beta_d = \alpha \gamma + \beta$. 
\end{enumerate}

As in the naive regression case, we equate $\alpha = \beta_1$ and omit it from the prior evaluations. Again, in this paper we use independent shrinkage priors (\ref{horseshoe}) over all the elements of $\beta_c$ and $\beta_d$.

Finally, to improve mixing over the parameter of interest, $\alpha$, we add an additional step of sampling $\alpha \mid \beta_c, \beta_d, \sigma_{\epsilon}, \sigma_{\nu}$. With a flat prior over $\alpha$ this step amounts to a simple linear regression update with response $\tilde{Y} = Y - \mX \beta_d$ and predictor variable $\tilde{Z} = Z - \mX \beta_c$, which is a normal distribution with mean $(\tilde{Z}^t\tilde{Z})^{-1}\tilde{Z}^t\tilde{Y}$ and variance $\sigma_{\nu}^2 (\tilde{Z}^t\tilde{Z})^{-1}$. Note that this step is not possible in the naive parametrization because in that case $\alpha$ cannot be updated separately from $\beta_d$.

\end{appendix}
\end{document}